\documentclass[useAMS,usenatbib]{mn2e}

\usepackage{graphicx}
\usepackage{times}
\usepackage{natbib}
\usepackage{color}
\usepackage{ulem}

\usepackage{amsmath}
\usepackage{amssymb}
\usepackage{float}
\usepackage{todonotes}

\newcommand{\apjs}{ApJS}

\newcommand{\apjl}{ApJL}

\newcommand{\aap}{A \& A}

\newcommand{\enzo}{\it{\small ENZO}}

\begin{document}
 
\title[Evolution of magnetic fields in the ICM] {Dynamical evolution of magnetic fields in the intracluster medium}
\author[P. Dom\'inguez-Fern\'andez et al.]{P. Dom\'inguez-Fern\'andez$^{1}$\thanks{E-mail: pdominguez@hs.uni-hamburg.de   } , F. Vazza$^{2,3,1}$ , M. Br{\"u}ggen$^1$ and G. Brunetti$^3$\\
$^{1}$ Hamburger Sternwarte, Gojenbergsweg 112, 21029 Hamburg, Germany\\
$^2$  Dipartimento di Fisica e Astronomia, Universit\'{a} di Bologna, Via Gobetti 92/3, 40121, Bologna, Italy\\
$^{3}$  Istituto di Radio Astronomia, INAF, Via Gobetti 101, 40121 Bologna, Italy}

\date{Received / Accepted}
\maketitle
\begin{abstract}
We investigate the evolution of magnetic fields in galaxy clusters starting from constant primordial fields using highly resolved ($\approx \rm 4 ~kpc$) cosmological MHD simulations. The magnetic fields in our sample exhibit amplification
via a small-scale dynamo and compression during structure formation.
In particular, we study how the spectral properties of magnetic fields are affected by mergers, and we relate the measured magnetic energy spectra  to the dynamical evolution of the intracluster medium. 
The magnetic energy grows by a factor of $\sim$ 40-50 in a time-span of $\sim 9$ Gyr
and equipartition between kinetic and magnetic energy occurs on
a range of scales ($< 160 \rm ~kpc$ at all epochs) depending on the
turbulence state of the system.
We also find that, in general, the outer scale of the magnetic field and the MHD scale  are not simply correlated in time. 
The effect of major mergers is to shift the peak magnetic spectra to {\it smaller scales}, whereas the magnetic amplification only starts after $\lesssim$ 1 Gyr.
In contrast, continuous minor mergers promote the steady growth of the magnetic field. We discuss the implications of these findings in the interpretation of future radio observations of galaxy clusters. 
\end{abstract}

\label{firstpage} 

\section{Introduction}
\label{sec:intro}

Galaxy clusters assemble through mergers and
accretion until they reach an approximate virial equilibrium \citep[e.g.][]{2012ARA&A..50..353K,2015SSRv..188...93P}. 
These events affect the
 space between galaxies which is filled with a dilute plasma, 
 known as the intracluster medium (ICM). In particular, radio observations shed light on the non-thermal
component of the ICM revealing the existence of cosmic rays and
magnetic fields permeating galaxy clusters
 \citep[e.g.][]{2012A&ARv..20...54F,2014IJMPD..2330007B,review_dynamo}.
Observations of synchrotron emission indicate magnetic fields with strengths of a few $\mu G$  (corresponding to an energy density of $\sim 1-2 \%$ of the thermal energy of the ICM) and 
typical coherence scales in the range of $\sim 10-50$ kpc \citep[e.g.][]{2005A&A...434...67V}.
Typically, this coherence scale is derived by a Fourier analysis of rotation measure (RM) maps, and inferring 
the maximum and minimum scales in the magnetic spectrum (often assuming a Kolmogorov-like power spectrum) necessary to reproduce the observed properties within uncertainties \citep[e.g.][]{mu04,bo10,bo13}.
In order to explain their observed morphology and strength, it has been suggested that magnetic fields get tangled over time by some other process than gas compression  \citep[e.g.][]{do99,br05,xu09}.

While the origin of magnetic fields in galaxy clusters
is still subject to debate, two scenarios have been widely discussed: (i) the 
\textit{primordial} scenario, in which 
magnetic fields have been generated in the early Universe
possibly during (or after) inflation but prior to the formation 
of galaxies 
\citep[e.g.][]{PhysRevD.37.2743,1475-7516-2014-05-040,
2001PhR...348..163G,
2011PhR...505....1K,
2016RPPh...79g6901S}
 and (ii) the \textit{astrophysical} scenario,
in which magnetic fields were produced 
from stellar winds 
\citep[e.g.][]{2009MNRAS.392.1008D} or active galactic nuclei (AGN) \citep[e.g.][]{2011ApJ...739...77X}.
A lower bound on the strength of the initial seed field of $B\geq 3 \times 10^{-16}$G (comoving) has been inferred for voids from the non-observation of secondary gamma-rays around blazars \citep[e.g][]{2010Sci...328...73N}. 
On the other extreme,
upper limits of the order of $B\leq 10^{-9}$G (comoving), derived from the observed level of 
cosmic microwave background (CMB) anisotropies 
\citep[e.g.][]{2016A&A...594A..19P}, can be used to limit the 
strength of any primordial seed field with coherence scales of $\sim$ Mpc or larger. 

Regardless of the magnetogenesis scenario, magnetic fields must have been significantly
amplified in order to have reached today's values. 
It is generally assumed that the amplification of the initial magnetic fields occurred via the combined effect of adiabatic compression and the presence of a small-scale dynamo, both of which are driven by minor or major mergers
\citep[e.g.][]{1999ApJ...518..594R,2005ApJ...631L..21B,2016RPPh...79g6901S}. 
The presence of a small-scale dynamo requires the existence of turbulence in the ICM, which is supported by cosmological simulations \citep[e.g.][]{do05,va09turbo,in08,ry08,lau09,va11turbo,2015MNRAS.453.3999M,2018MNRAS.480.5113M,review_dynamo} and more recently, also by observations \citep[e.g.][]{Hitomi2017}. 
A dynamo process converts kinetic energy into
magnetic energy over the typical dynamical timescales of the turbulent cascade. 
It is believed that the amplification of ICM magnetic fields arises 
from the turbulence developing on scales which are a fraction of cluster virial radius ($\leq 0.5-1 ~\rm Mpc$) \citep[e.g.][and references therein]{review_dynamo}. Previous simulations
have shown that only a few percent of the incompressible turbulent energy needs to be dissipated to account for the observed field strength \citep[e.g][and references therein]{2015Natur.523...59M}.

Whenever the characteristic scale of the magnetic field is comparable or smaller than the characteristic scale of
fluid motions, the dynamo is referred to as a \textit{small-scale dynamo} (also called fluctuation dynamo)
\citep[e.g][]{1983flma....3.....Z,1967Kasan}. Conversely, a \textit{large-scale dynamo} refers to magnetic fields that
are spatially coherent on scales comparable to the scale of the underlying
astrophysical system \citep[e.g][]{1983flma....3.....Z,1978mfge.book.....M}.
Since galaxy clusters do not show substantial rotation,
it is likely that
the turbulent small-scale dynamo
winds up magnetic fields on scales smaller than the turbulence injection scale
\citep[e.g][]{2006MNRAS.366.1437S,2012SSRv..169..123B,1967Kasan,1967PhFl...10..859K,1992ApJ...396..606K,2007mhet.book...85S,2006ApJ...640L.175B,2008PhRvL.100h1301S}. 

In previous papers 
\citep[e.g][]{2016ApJ...817..127B,2015Natur.523...59M},
 driven turbulence in the ICM has been studied in
a cosmological context. Still, it remains a challenge to push the spatial resolution down to the so-called \textit{MHD scale} ($l_A$) at which the magnetic energy is strong enough to  prevent additional bending of the magnetic field lines. 
It is crucial to resolve $l_A$ in order to fully capture the development 
of the small-scale dynamo amplification, but  $l_A$ can in principle be extremely small ($\ll \rm kpc$) for arbitrarily small seed magnetic fields. The Reynolds number achieved in simulations is also an important factor that directly affects the magnetic field growth.
While the Reynolds number based on the full \textit{Spitzer} viscosity in the ICM is believed to be of the order of $R_e \sim 10^2$ \citep[e.g][]{brunetti2007,2014ApJ...797..133C}, the reduced proton mean free path in the collisionless ICM can result in much larger Reynolds numbers \citep[][]{2016ApJ...817..127B,2011MNRAS.412..817B}. 
This suggests that the fluid approximation provides a suited model for the properties of the ICM \citep[e.g][]{2017MNRAS.465.4866S,2014ApJ...781...84S}.

More recently, it has been shown that initial magnetic
field seeds can be amplified via a dynamo up to strengths of $\sim\mu$G in
cosmological grid simulations
\citep[e.g][]{va18mhd} 
(hereafter Paper I). Here, we present a new sample of galaxy clusters to
study the spectral properties of each galaxy cluster in our sample.
Firstly, we study the characteristic spectral features
of the magnetic energy in different types of clusters at $z=0$.
Secondly, we follow the spectral evolution of a particular
cluster that is merging. 

The paper is structured as follows: in Section \ref{section:methods} we 
present the numerical setup and describe the fitting process of the magnetic energy spectra. In Section \ref{sec:res} we present our results in two parts, the
first one dedicated to the properties of our galaxy cluster sample at $z=0$, and the second one
describing the evolution of a merging cluster. In Section \ref{section:numerics}
we discuss numerical aspects and in Section \ref{sec:conclusions},
we discuss the implications of our results.

\section{Methods}\label{section:methods}

\subsection{The Simulated Dataset}

We simulated the formation of massive galaxy clusters in a cosmological framework with the {\enzo} grid code \citep[][]{enzo13}.  We  used the Dedner formulation of MHD equations \citep[][]{wa09} and used adaptive mesh refinement (AMR) to  increase the dynamical resolution within our clusters, as in Paper I.
We assumed a $\Lambda$CDM cosmology ($h = 0.72$, $\Omega_{\mathrm{M}} = 0.258$, $\Omega_{\mathrm{b}}=0.0441$ and  $\Omega_{\Lambda} = 0.742$) as in \citet[][]{va10kp}.

Each cluster was selected in a comoving volume of (260 Mpc)$^3$, first simulated at coarse resolution \citep[][]{va10kp}, and then resimulated with nested initial conditions \citep[][]{2007ApJ...665..899W}. We employed two levels of static uniform grids with $256^3$ cells each and using $256^3$ particles each to sample the dark matter distribution, with a mass resolution per particle of $m_{\rm DM}=1.3 \cdot 10^{10} M_{\odot}$ at the highest level. 

Then, we further refined the innermost  $\sim$ (25 Mpc)$^3$ volume, where each cluster forms, with additional 7 AMR levels (refinement $=2^7$).  The refinement was initiated wherever the gas density was $\geq 1\%$ higher than its surroundings. This gives us a maximum spatial resolution of $\Delta x_{\rm max}=3.95 ~\rm kpc$ per cell.

With our setup (see Paper I), for  $z\leq 1$ the virial  volume of clusters is 
refined at least up to the 6th AMR level ($15.8$ kpc) at $z=0$, and most of the central volume within $\leq 1$ Mpc from the cluster centre is simulated with 3.95 kpc/cell.\footnote{Each cluster simulation used $\sim 30,000-50,000$ core hours running on 64  nodes on JUWELS at J\"ulich Supercomputing Centre.}

In this work, we will only discuss  {\it non-radiative} cosmological simulations, meaning that we only included  the effect of cosmic expansion, gas, Dark Matter self-gravity and (magneto)hydrodynamics, in order to solely focus on the growth of magnetic fields by the turbulence induced by structure formation.

In order to seed magnetic fields at the beginning of our runs, we mimic a simple
primordial origin of magnetic fields, in which we  initialized the field to a uniform value $B_0$ across the entire computational domain, along each coordinate axis. The initial magnetic seed field of $0.1 ~\rm nG$ (comoving) is chosen to be below the upper limits from the analysis of the CMB \citep[e.g.][]{sub15}. This particular setup is easy to implement, ensures $\nabla \cdot \vec{B}=0$ by construction, and has been already tested in our previous work on the subject \citep[][]{va14mhd,va18mhd}. Moreover, several studies have shown that the impact on the initial magnetic field topology within galaxy clusters (provided that the simulated dynamical range is large enough to enter the dynamo regime) is negligible \citep[e.g.][]{2015MNRAS.453.3999M,va17cqg,va18mhd}, hence our results do not strongly depend on this particular setup. \\
We refer the reader to Appendix \ref{appen_2} for a short overview of the key findings of Paper I. There, we showed that our numerical setup provides enough resolution to resolve the MHD scale, $l_A$, in a large fraction of the cluster volume  during its late evolution ($z \leq 1$). Moreover, the simulations show features of   small-scale dynamo amplification. However, as we discuss in depth in Sec.~\ref{section:numerics}, 
some results can be affected by the limited spatial resolution.

\subsection{Fitting the magnetic power spectrum}
\label{section:fit}

The three-dimensional power spectrum is defined as
\begin{equation}
P_{ij} (\mathbf{k})= \frac{1}{(2\pi)^3}\int \int \int 
e^{-i\mathbf{k} \cdot \mathbf{x}}
R_{ij} (\mathbf{k}) d\mathbf{k},
\end{equation}
where 
$R_{ij}=\left\langle u_i(\mathbf{x_0})u_j(\mathbf{x_0} + \mathbf{x}) \right\rangle$ 
is the two-point correlation function between the velocities $u_i$ and $u_j$
\citep[e.g.][]{batchelor_1951}.
When the corresponding fields do not depend on the position and only 
depend on the distance between two points, i.e. 
we consider homogeneous
and isotropic fields, the total energy is given by
\begin{equation}\label{spec_energy}
E_{tot} = \frac{1}{2}  \left\langle u_i^2 \right\rangle = 
\frac{1}{2} R_{ii}(\mathbf{0})=
\int_{0}^{\infty} E(k) dk,
\end{equation}
where $E(k)$ is thus the scalar energy distribution per unit mass for the
mode $k$ related to the diagonal components of the tensor $R_{ij}$, and
therefore, the relation between this spectral energy and the 
one-dimensional power spectrum is found to be
\begin{equation}
E(k)=2\pi k^2P_{ii}(k).
\end{equation}
This approximation works well for the rather chaotic and isotropic velocity
field always found in cosmological cluster simulations \citep[e.g.][]{do05,va11turbo,wi17}.
We computed first the power spectrum
by using standard algorithms for the
three-dimensional Fast Fourier Transform (FFT)
of the velocity and magnetic fields within the simulation box and
then by summing up the contributions over spheres within a radius
$k=\sqrt{k_x^2+k_y^2+k_z^2 }$ in Fourier space. Finally, by multiplying
by the factor $2\pi k^2$, we obtained the energy spectrum of the
magnetic and velocity field. 

\begin{figure}
    \centering
    \includegraphics[width=9.5cm]{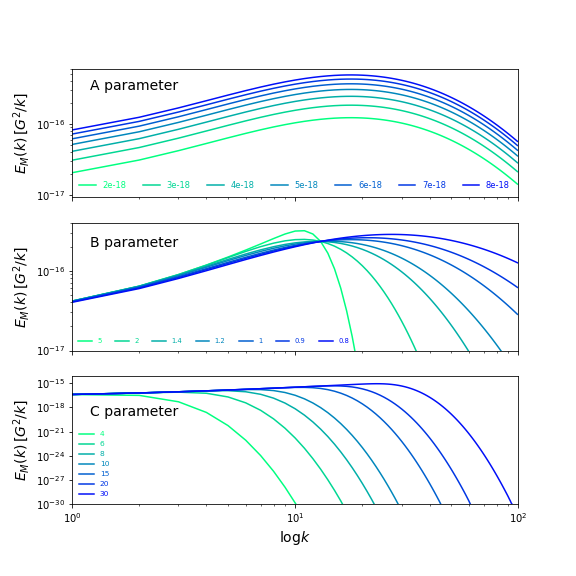}
    \caption{Variation of A, B and C parameters in Eq. (\ref{fit_eq}). \textit{Top panel:} change in the normalization. \textit{Middle panel:} change in the width. \textit{Bottom panel:} change in the position of the outer scale.}
    \label{fig:param}
\end{figure}

While the velocity power spectra can be characterized by a power-law and by an injection scale, the magnetic spectra are more complex. We fit the magnetic spectra by the equation:
\begin{equation}\label{fit_eq}
E_M(k)= A \,k^{3/2}
\left[ 1- \text{erf} \left[ B 
\ln\left( \frac{k}{C} \right)  \right]  \right],
\end{equation}

where the $A$ parameter gives the normalization of the magnetic spectrum, $B$ is related to the width of the spectra 
and $C$  is a characteristic wavenumber corresponding 
to the inverse outer scale of the magnetic field (see Fig. \ref{fig:param}).  Eq. (\ref{fit_eq}) is rooted in dynamo theory as a solution for single-scale turbulent flows (\citet{1967Kasan}, \citet{1967PhFl...10..859K}, \citet{1992ApJ...396..606K}).
In the remainder of the paper we propose to use  Eq. (\ref{fit_eq}) as a proxy to characterize our evolving magnetic spectra with a minimal set of parameters ($A$, $B$, $C$ as detailed above), even though the equation is not valid for the scales and conditions that we are studying.
It should be stressed that the aim of the paper is not to connect directly these parameters with Kazantzev's dynamo model since the generation and evolution of 
turbulent magnetic fields in the ICM are affected by a hierarchy of complex processes. In particular, we note that:

1) The assumptions under which Eq. (\ref{fit_eq}) is derived, such as having a single-scale turbulent flow, a Kolmogorov spectrum for the velocity field, neglecting the resistive scale, etc. 
(see more details on the assumptions and derivation in \citealt{1992ApJ...396..606K}) are not valid since, in our system, laminar gas motions and advection at many scales 
may also affect the topology of the magnetic fields in the ICM. Furthermore, the  magnetic
field is amplified and re-shaped by the turbulence generated every time a merger occurs.

2) The analysis of non-linear effects such as ambipolar diffusion or magnetic reconnection are far beyond the scope of this work. But we can comment that some of these affects have been also studied in \citet{1992ApJ...396..606K}, where the final magnetic power spectrum exhibits a similar shape, i.e. a power law multiplied by a Macdonald function (or modified Bessel function of second order) of different orders. For small $k$, they can reduce to
Eq. (\ref{fit_eq}).

3) As long as the velocity scale responsible for the dynamo forcing is larger than the scales where the magnetic energy spectrum peaks, Eq. (\ref{fit_eq}) is valid.  This condition is matched during the initial stage of cluster formation, and is later violated after the magnetic field has grown to larger scales.  It is our intention to quantify the development of magnetic fields as a function of resolution (as in Paper I) as well as of the cluster evolution. For this reason, it is convenient to apply Eq. (\ref{fit_eq}), as the dynamo in our runs is expected to stay in the kinematic regime for long due to the finite numerical resolution \citep[e.g.][]{2016ApJ...817..127B}.

\section{Results}
\label{sec:res}

\subsection{Magnetic fields in the cluster sample}
\label{subsec:cluster_sample}

\begin{figure*}
	\includegraphics[width=\textwidth]{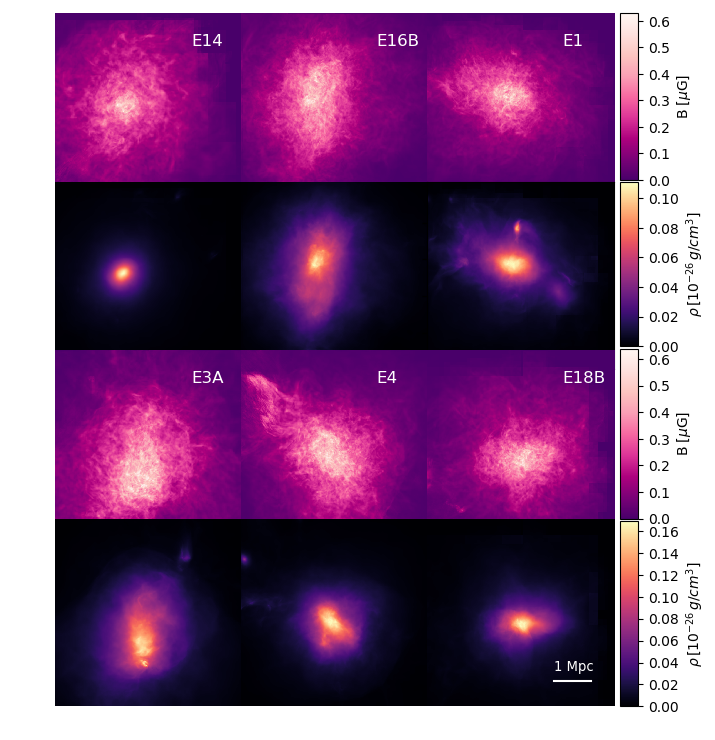}
    \caption{Maps of projected gas density and magnetic field strength for all clusters in our sample at $z=0$ (we omit cluster E5A since this cluster is analyzed in detail in Section \ref{section:E5A_evolution} and \ref{sec:spectral_evolution}). The
    main characteristics of these clusters can be found in Tab. 
    \ref{tab:tab1}.}
    \label{fig01:Maps_allz0}
\end{figure*}

In this section we analyze a sample of seven clusters in different
dynamical states: clusters with ongoing mergers (ME) at $z=0$,
relaxed ones (RE) and post major merger ones
(PM). 
In Fig. \ref{fig01:Maps_allz0} 
we show the projected gas density and magnetic
field strength for all of our clusters, considering the highest resolution of
our simulation (3.95 kpc). 

A list of the main parameters of our simulated clusters is given in Tab.~\ref{tab:tab1}. The estimate of the total (gas+DM) mass inside $R_{\rm 100}$, as well as the tentative classification of the dynamical state at $z=0$ of each object follows from \citet{va10kp}.
Our dynamical classification is done in two steps:
firstly, clusters with a major merger (based on the total mass accretion 
history within $R_{\rm 100}$) for $z<1$ are 
classified as \textit{post-mergers} (PM). 
In particular, major mergers in the range $0 \leq z \leq 1$ are selected 
considering that the change of the mass increment $\xi=M_2/M_1$ is 
 $\xi>0.3$, where $M_1$ is the mass at a time $t$ and $M_2$ is the mass at time $t+1$ Gyr   
 \citep[][]{2010MNRAS.406.2267F}. 
Secondly, if no major merger is found in this time interval, 
we additionally compute the ratio 
between the total kinetic energy of gas motions inside $R_{\rm 100}$, $E_{K}$, and the total energy ($E_{\rm tot}=E_{K}+E_{T}$)
inside the same volume. 
This parameter has been shown to characterise the dynamical activity of clusters well \citep[e.g.][]{TO97.2}. {\it Relaxed}
(RE)
 clusters typically have $E_{K}/E_{\rm tot}<0.5$ while 
 {\it merging} (ME) clusters have  $E_{K}/E_{\rm tot} \geq 0.5$. 
 In Tab.~\ref{tab:tab1} we also list the redshift of the 
 last major merger ($z_{\rm last}$) for post-merger systems, while for 
 relaxed systems we conventionally consider $z_{\rm last}=0$ 
 and $z_{\rm last}=1$ for merging systems. 
 For a more detailed discussion of the classification scheme 
 we refer the reader to \citet{va10kp} and references therein. 

\begin{table*}
\centering \tabcolsep 5pt 
\begin{tabular}{c|c|c|c|c|c|c|c}
  ID & $M_{\rm 100} \rm [M_{\odot}]$ & $R_{\rm 100}$ [Mpc]  & Dynamical state & $B_0 [\rm \mu G]$ & A ($10^{-17} \, [\textnormal{G}^2/k]$) & B ([-]) & C (k[1/2 Mpc]) \\ \hline
  E14 & $1.00  \cdot 10^{15}$  &  $2.60$   & RE& 1.726 &5.470 $\pm$  0.111 
& 1.090 $\pm$ 0.009
& 6.461 $\pm$ 0.096 \\
E5A & $0.66  \cdot 10^{15} $ & $2.18$   & ME& 1.050 &1.985 $\pm$ 0.059 
& 1.054 $\pm$ 0.012
& 8.708 $\pm$ 0.192 \\
  
  E1 & $1.12 \cdot 10^{15}$ & $2.67$ &  PM $(z_{\rm last}=0.1)$& 1.308 & 2.052  $\pm$  0.036 
& 1.118 $\pm$ 0.009 & 10.052 $\pm$ 0.131 \\
  E3A & $1.38 \cdot 10^{15} $ & $2.82$  & PM $(z_{\rm last}=0.2)$& 1.672 &2.372  $\pm$ 0.041 
& 1.167 $\pm$ 0.009
& 8.936 $\pm$ 0.110 \\
  E16B & $1.90 \cdot 10^{15}$ & $3.14$  & PM $(z_{\rm last}=0.2)$& 2.474 &9.041 $\pm$ 0.164 
& 1.134 $\pm$ 0.009
& 10.437 $\pm$ 0.138 \\
  E4 & $1.36  \cdot 10^{15}$ & $ 2.80$  & PM $(z_{\rm last}=0.4)$ & 1.572 &4.521 $\pm$ 0.074 
& 1.124 $\pm$ 0.008
& 10.236 $\pm$ 0.123 \\
  E18B & $1.37  \cdot 10^{15}$ & $2.80$ & PM $(z_{\rm last}=0.5)$& 1.716 &3.396 $\pm$ 0.049 
& 1.113 $\pm$ 0.007
& 9.974 $\pm$ 0.106 \\
  \end{tabular}
  \caption{Main parameters at $z=0$ of the galaxy clusters analyzed in this work. The 4th column lists the tentative dynamical classification of each object (with the approximate redshift of the last major merger, in the case of post-merger clusters). The value of $B_0$ is the mean magnetic field within
  200 kpc from the corresponding radial profiles plotted in Fig. \ref{fig:B_profiles}}
  \label{tab:tab1}
\end{table*}

\begin{figure}
    \centering
    \includegraphics[width=\columnwidth]{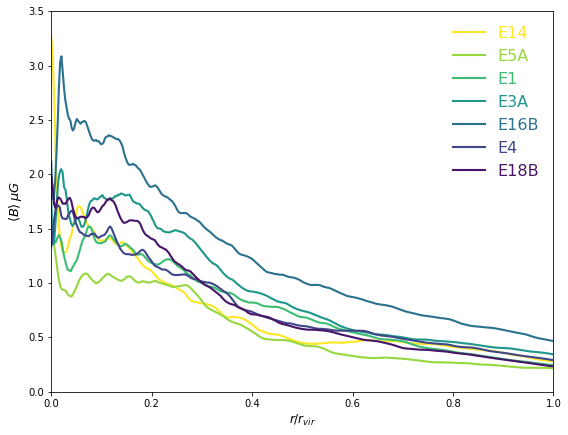}
    \caption{Radial profile of the magnetic field for all the clusters in our sample
    at $z=0$ computed at our highest resolution run at the 8th AMR level.}
    \label{fig:B_profiles}
\end{figure}

For each cluster, we computed the radial profile of the average magnetic field from the peak of gas density at $z=0$ at the highest resolution,  as shown in Fig. \ref{fig:B_profiles}.
Within the sample variance, we find that the magnetic field follows gas density
as $B(n) \approx B_0 \cdot (n/n_0)^{0.5}$ (where $n$ is the gas density and $n_0$ is the core gas density) as in Paper I. 
In fact, the radial profiles in Paper I appear to 
be consistent with what can be derived by Faraday Rotation analysis of the Coma cluster \citep[][]{bo13}, despite the fact that the distribution of magnetic field components found in our simulations deviate significantly from a Gaussian distribution.
The central magnetic field value of each cluster, $B_0$, is given for reference in Tab.~\ref{tab:tab1}. In general, we can see that the most perturbed cluster (E5A) does not show the strongest fields. The
central value, $B_0$ (measured as the average within the innermost $\leq 200 \rm ~kpc$ radius from the cluster centre), is strongly correlated with the mass of the cluster. Indeed, in
Fig.~\ref{fig:B_profiles} we can see that the higher the mass, the higher the central
value of the magnetic field.  While observations do not show a clear correlation of the mean magnetic field with the mass of the host cluster \citep[e.g.][]{2017A&A...603A.122G}, our normalization $A$, which is the parameter most closely linked to the Faraday Rotation, shows little correlation with mass, and has a large scatter.

\subsection{Spectral properties in the cluster sample}
\label{sec:spectral_sample}

\begin{figure}
	\includegraphics[width=\columnwidth]{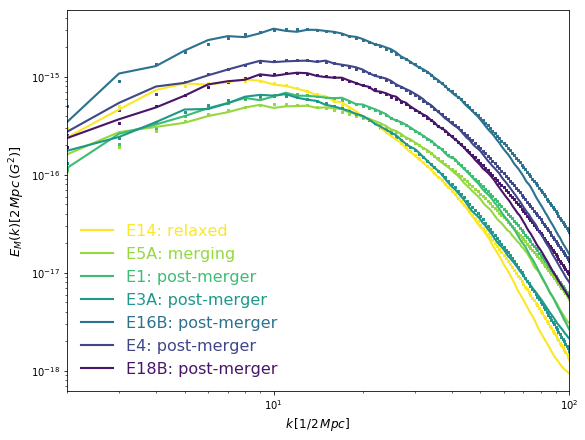}\\
	\includegraphics[width=\columnwidth]{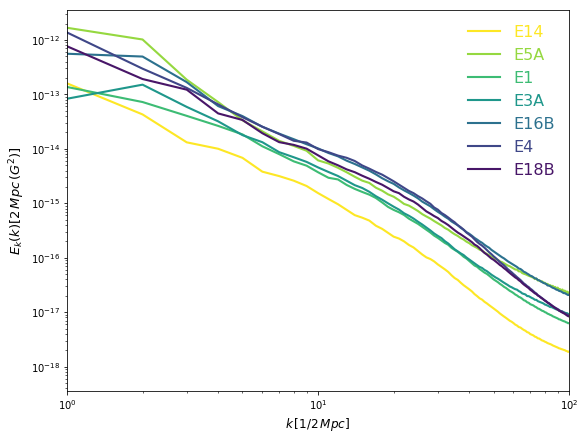}\\
	\includegraphics[width=\columnwidth]{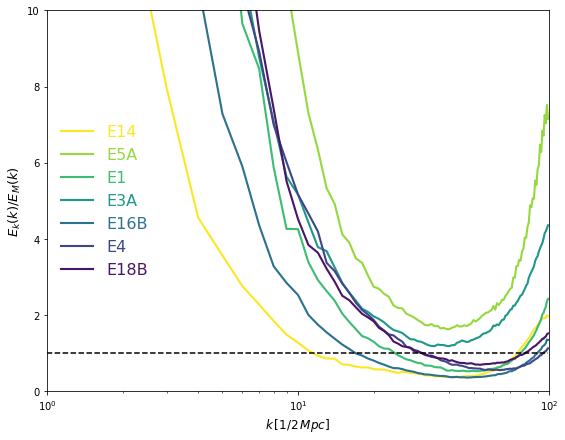}
    \caption{Magnetic energy (\textit{top panel}) and kinetic energy (\textit{middle panel}) spectra of all of our cluster sample at $z=0$. The kinetic spectra were multiplied by $\sqrt{n}$, where
    $n$ is the gas density, in order for the spectra in both panels to have the same units.
    The solid lines
    correspond to the data and the scatter plots show the best-fit of the corresponding data
    using Eq. (\ref{fit_eq}). In the \textit{bottom panel} we show the ratio of kinetic to magnetic energy, $E_K/E_M(k)$, the horizontal black dashed line indicates where we have equipartition.}
    \label{fig0:Spectrum_all}
\end{figure}

Next, we proceeded to compute the magnetic energy power spectra for the innermost region of all clusters at $z=0$ as described in Sec.~\ref{section:fit}
. We computed power spectra only for the innermost $\approx 2^3 ~\rm Mpc^3$ region of each cluster, where the resolution is approximately constant and equal to the 8th and maximum AMR level (corresponding to a $512^3$ grid).  By doing so, we can neglect the effect of coarse-mesh
effects in our FFT analysis as the majority of the central cluster volume is refined up to the highest level for all our clusters \citep[see discussion in ][]{va18mhd}.
 The corresponding spectra, along with the best-fit curves are plotted 
in the top panel of Fig. \ref{fig0:Spectrum_all} and the best-fit parameters are
listed in Tab. \ref{tab:tab1}. To a good degree of approximation, 
all spectra are well fitted 
by Eq. (\ref{fit_eq}) regardless of the dynamical state of each cluster. All clusters in the sample show similar spectral shapes, with a peak of magnetic energy in the range $\sim 200-300 ~\rm kpc$ and differences in normalization of a factor $\leq 5$. 
As shown in \citet{2018MNRAS.474.1672V}, this non-Gaussian distribution of magnetic field strengths may result from the superposition of multiple magnetic field components that have been accreted at different times via mergers.
For completeness, we also show the 
kinetic spectra of all the clusters in the central panel of 
Fig.~\ref{fig0:Spectrum_all}. These kinetic spectra
are very similar, i.e. we observe a higher normalization for perturbed clusters as there is more
turbulence involved in these systems, and the lowest normalization is observed for the relaxed
cluster (E14). Comparing this to the magnetic spectra shown in the 
top panel of Fig.~\ref{fig0:Spectrum_all}, we can clearly see that a
higher level of turbulence does
not necessarily imply higher values of the magnetic field. 
This may seem counter-intuitive but it is caused by the fact that the amplification of magnetic fields from small to large spatial scales is a slow process that takes a few eddy turnover times. 
Therefore, even in the presence of a large input of turbulent kinetic energy, significant magnetic 
amplification can only be observed with a delay of $\sim \rm ~Gyr$. While part of this delay is caused by numerical effects (e.g. our numerical finite growth rate depends on the limited Reynolds number our simulation can resolve), this delay is of the same order as the eddy turnover timescale for $\sim 500 ~\rm kpc$ turbulent eddies being injected with a $\sigma_v \sim 500 ~\rm km/s$ velocity. This is the necessary time span for turbulence to cascade down to the scales that 
can drive a dynamo growth. 

In the bottom panel of Fig.~\ref{fig0:Spectrum_all} we plot the ratio between kinetic 
and magnetic energy in order to visualize the scales at which equipartition is reached.
RE systems reach equipartition at larger scales compared to PM systems, which is consistent
with the general picture of a small-scale dynamo acting according to the amount of turbulence in the system. 
As expected, we also observe that the ME system is still not in equipartition at larger scales because this is the
most perturbed cluster and it is mostly dominated by compressive turbulence.
%

\subsubsection{Parameterization of cluster magnetic spectra}

 Our analysis in Paper I supports that 
the magnetic spectra show signs of a dynamo near saturation (see Appendix \ref{appen_2}).  
However, as we shall see in Sec.~\ref{section:E5A_evolution}, 
if a small-scale dynamo is acting, it co-exists with bulk motions on larger scales that are affecting the evolution of the magnetic field during the whole assembly history of the clusters. As a consequence, the magnetic properties in our sample result from the cumulative (and discontinuous)
action of dynamo during the entire cluster life-time. Therefore, there is no immediate connection between the spectral magnetic properties and the turbulent properties of the cluster at a given time.

In order to study how the best-fit parameters, $A$, $B$ and $C$,
are related to the mass, dynamical state and redshift since the last major merger, we produced 
Figs.~\ref{fig0:Params_vs_M}, \ref{fig0:Params_vs_z} and \ref{fig0:AC_vs_M}. For our limited sample, we can conclude:

\begin{figure}
	\includegraphics[width=\columnwidth]{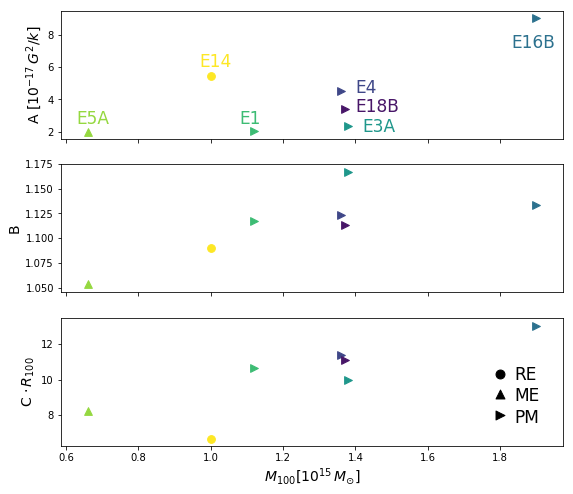}
    \caption{Comparison of best-fit parameters of each cluster in our sample at $z=0$ 
     according to their virial mass.}
    \label{fig0:Params_vs_M}
\end{figure}
\begin{figure}
	\includegraphics[width=\columnwidth]{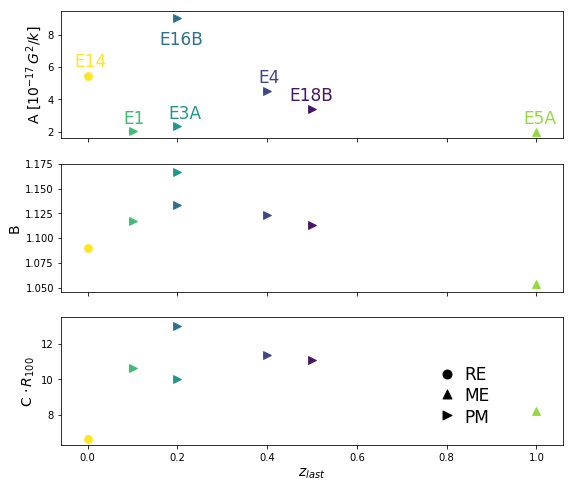}
    \caption{Comparison of best-fit parameters of each cluster in our sample to their last major merger event.}
    \label{fig0:Params_vs_z}
\end{figure}
\begin{figure}
	\includegraphics[width=\columnwidth]{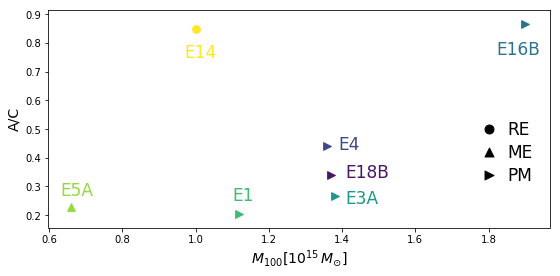}
    \caption{Combination of best-fit parameters of each cluster (proportional to the rotation measure (RM)) in our sample compared to their virial mass.}
    \label{fig0:AC_vs_M}
\end{figure}

\begin{itemize}
        \item[1)] The spectrum normalization ($A$): We find a dependence of the mass of the host cluster, and also a hint of a dependence on the dynamical
        state of the cluster. For a given mass bin we find $A_{\rm{ME}}<A_{\rm{PM}}<A_{\rm{RE}}$.  On the other hand, we do not find a correlation with the time since the 
        last major merger. 
       
        \item[2)]  The spectrum width ($B$): 
        This parameter is found to depend on the dynamical state of each cluster, i.e.  $B$ is larger in
        less perturbed systems (PM) and smaller in highly
        perturbed systems (ME). This presumably means that relaxed
        systems have had enough time for 
        turbulence to cascade to the small scales and amplify the magnetic field during past mergers
        resulting in a broader spectrum; whereas the merging systems have a more restricted region of magnetic amplification. 
        As will be mentioned in \ref{sec:spectral_evolution}, mergers shift the spectrum towards smaller scales, thus the combination of this shift and the narrow spectrum means that merging systems contain more small-scale eddies. Therefore perturbed systems have a higher magnetic growth rate than relaxed systems.

        \item[3)] The inverse of the outer scale of magnetic field ($C$): we find a hint of a dependence on the mass and the dynamical state of the cluster, 
        $C_{\rm{RE}}<C_{\rm{ME}}<C_{\rm{PM}}$. 
        On the assumption of the existence of a small-scale dynamo,
        this would suggest that in more relaxed
        systems the dynamo had more time to grow towards larger scales (i.e. lower values of C).
         In the case of an on-going merger (as will be discussed in more detail in Sec.~\ref{section:E5A_evolution}), large scale gas motions may also
        affect the magnetic spectra merely by compression. In principle, it will be possible 
        that advection and buoyancy in a stratified medium such as the ICM also play a role in the outer scale behaviour, but the study of these effects are beyond the scope of this paper.
        \item[4)] Dependence on the time since the last major merger: 
        We did not find a correlation between the epoch of the last major merger ($z_{\rm last}$) and the best-fit parameters ($A$,$B$ and $C$), which suggests that the  
        magnetic energy spectrum at a given epoch does not retain much information about specific events, as well as that
        minor mergers are also an important player in setting the spectral properties of the ICM at $z=0$ (see discussion in \ref{sec:spectral_evolution}). Moreover, given the limited sample size we have here, it is difficult to disentangle effects connected to the mass of the host cluster and the dynamical state. 
        \item[5)] Rotation measure dependence on mass: The rotation measure (RM) of polarised radio emission from background sources scales as $|RM|  \propto  \int {B_{\rm ||}} \cdot n_e ~dl$, which can be approximated to $|RM| \propto B_{\rm \Lambda} \cdot \Lambda_B$, where $B_{\rm \Lambda}$ is the magnetic field at the autocorrelation scale $\Lambda_{B}$. To a first approximation, the total $|RM|$ from a clusters should scale with the $ \propto A/C $. 
        We find that the RM depends weakly on the mass(because of A and C dependence on mass), but scatters due to a dependence on the dynamical state of each cluster. This causes clusters with a factor $\sim 2$ difference in mass to show a very similar $A/C$ value (i.e. E14 and E16B, see Fig.\ref{fig0:AC_vs_M}).
      \end{itemize}

In order to link the evolution of magnetic spectra to the dynamical growth of galaxy clusters, in the next section we will analyze the assembly of one particular cluster.

\subsection{Detailed evolution of  cluster E5A}
\label{section:E5A_evolution}

We  studied the evolution of the cluster E5A by analysing a total of $\approx 100$ snapshots
in the range from $z=1.379$ to $z=0$. 
The cluster E5A is an interesting object as it forms via several
mergers in the course of nearly 9 Gyr.

\begin{figure*}
\includegraphics[width=\textwidth]{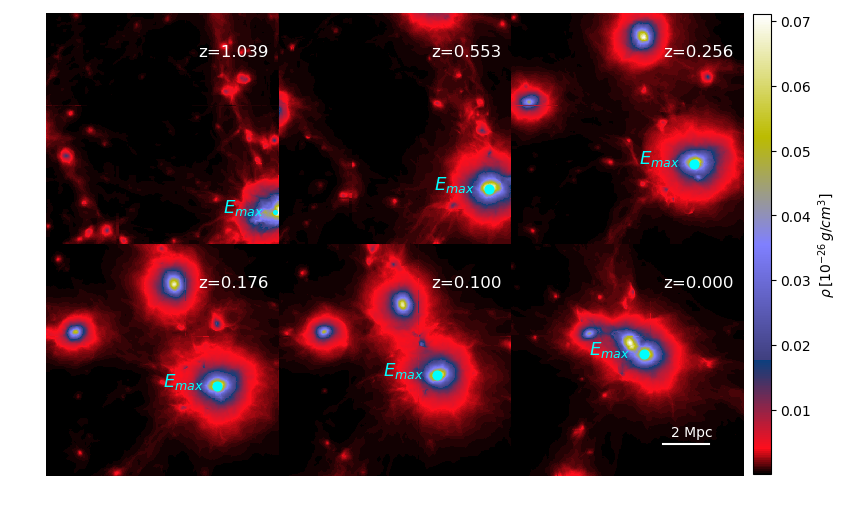} 
\includegraphics[width=\textwidth]{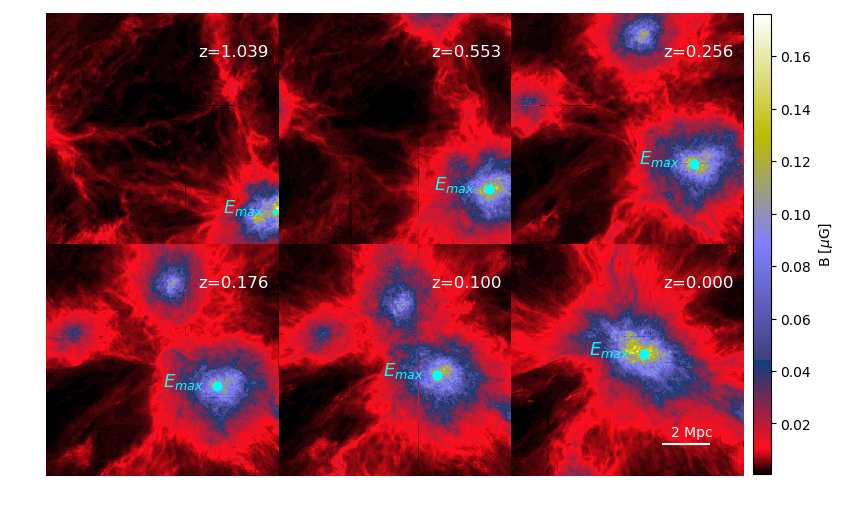}
\caption{\textit{Top panels:}
    projected z-component density field averaged a long the line-of-sight at various redshifts. \textit{Bottom
    panels:} projected z-component of the magnetic field strength averaged a long the line-of-sight at the same redshifts as the top panels. The dots indicate the centre of the
    most massive component at each redshift. The centre position was obtained by taking into account the kinetic energy within a simulation box of $640^3$ cells (see text for more details).}
    \label{Maps_seq}
\end{figure*}

\begin{figure*}
	\includegraphics[width=\textwidth]{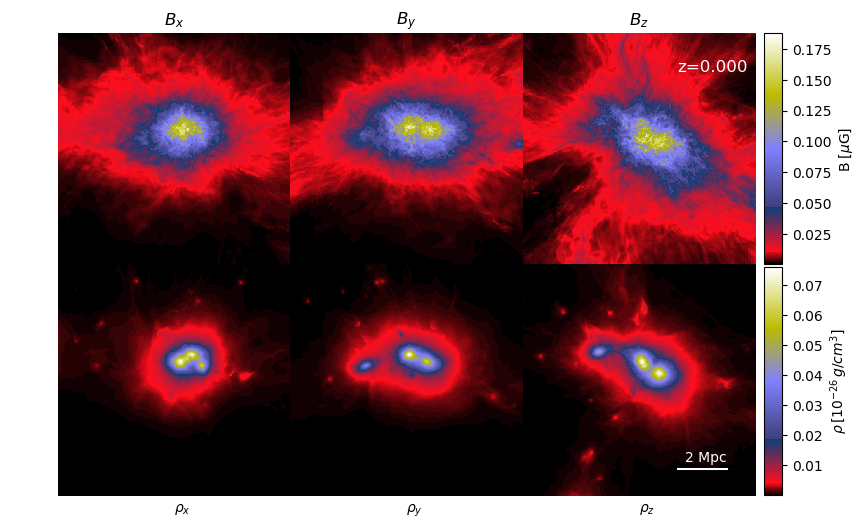}
    \caption{Maps averaged along the line-of-sight of the
    simulation box at z=0. The \textit{top panels} show the
    projected magnetic field strength and the \textit{bottom
    panels} show the projected density.}
    \label{fig01:Maps_z0}
\end{figure*}

We work on uniformly gridded data reconstructed at the
6th AMR level (15.8 kpc resolution). This is done even if the simulation has refined down to the 8th AMR level. In Fig. \ref{Maps_seq} 
we show snapshots of the
density and magnetic fields in a simulation box of $640^3$ cells. 
The maps in Figs.~\ref{Maps_seq}-\ref{fig01:Maps_z0} show a volume-weighted projection of the magnetic field strength along the line-of-sight in order to emphasize the diffuse magnetic field structure on large scales. Hence, the magnetic field values in the map are biased towards lower values than the ones measured in the computational box.

Next, we identified the centre of the main cluster (cyan dot in Fig. \ref{Maps_seq})
and then followed the evolution of gas and magnetic fields within a box of $100^3$ cells co-moving with the main cluster
centre. The trajectory of the centre was obtained by computing the 
location of the maximum of the thermal energy
after smoothing the data over a length of $\sim 20$ cells at each snapshot, and by applying a cubic spline time interpolation.

The final stage of the merger at $z=0$ is shown in Fig.~\ref{fig01:Maps_z0}, where
the volume distribution of the gas density and the magnetic 
field strength are plotted in \textit{x,y} and \textit{z} directions. 
The magnetic field distribution is asymmetric, showing a tangled structure and its
strength increases towards the centre of the major component. At least two prominent peaks in the magnetic field distribution (at the $\sim \rm \mu G$ level) near the central region are visible in all lines of sight. These peaks correspond to the largest and the second largest components. 

\begin{figure}
	\includegraphics[width=9.3cm]{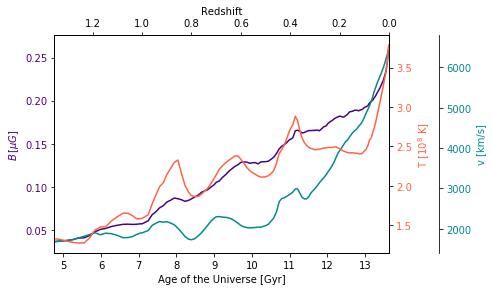}
    \caption{Evolution of the average magnetic field strength,    temperature and velocity of the $100^3$ simulation box}
    \label{fig02:Cluster_evolution}
\end{figure}

\begin{figure}
	\includegraphics[width=\columnwidth]{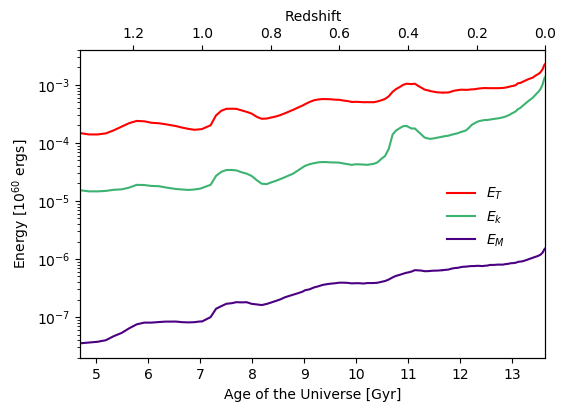}\\
	\includegraphics[width=\columnwidth]{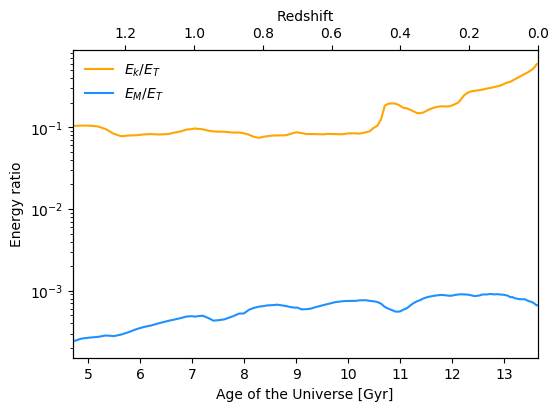}
    \caption{Energy evolution of the $100^3$ simulation box. The 
    \textit{top panel} shows the evolution of the thermal energy
    (red), kinetic energy (green) and magnetic energy (purple). The
    \textit{bottom panel} shows the corresponding energy ratios.}
    \label{fig03:Energy_evolution}
\end{figure}

In Fig.~\ref{fig02:Cluster_evolution} we show the evolution of
the magnetic field strength, temperature and velocity within
the moving simulation box. Every merger event is found to perturb the system and to increase the thermalization of the ICM, shown as 
peaks on the temperature evolution in Fig.\ref{fig02:Cluster_evolution}.  While gas velocity and gas temperature show pronounced peaks close each merger event, the
 evolution of the averaged magnetic
field strength is smoother.
 These smoothed peaks are correlated with the merger events, but show a delay of about $\sim 0.5$ Gyr with
respect to the velocity peaks.
In Fig.~\ref{fig03:Energy_evolution} we show 
the evolution of the total energy budget of the cluster (top panel)
and the corresponding evolution of the energy ratios
(bottom panel). 
During the whole period of evolution ($\sim 9$ Gyr), the magnetic energy has grown by a factor $\sim$ 40-50, the kinetic energy by a factor of $\sim$ 90-100, and the thermal energy has grown by a factor $\sim$ 10-20.
By the end of the simulation ($z=0$), the kinetic energy is $\sim$ 10-40\% of the thermal energy, while the magnetic energy is  $\sim 10^{-3}$ of the thermal energy.


\subsection{Spectral evolution of cluster E5A}
\label{sec:spectral_evolution}

\begin{figure}
	\includegraphics[width=\columnwidth]{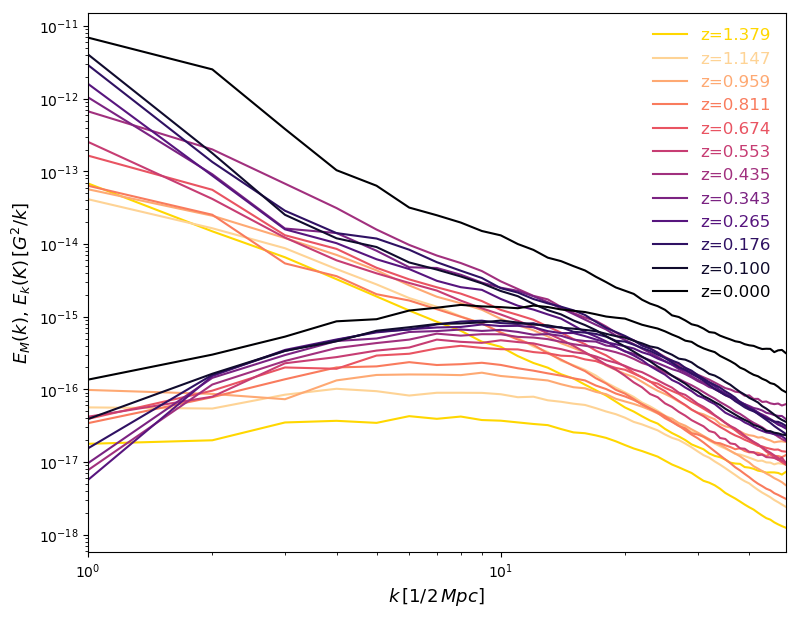}\\
	\includegraphics[width=\columnwidth]{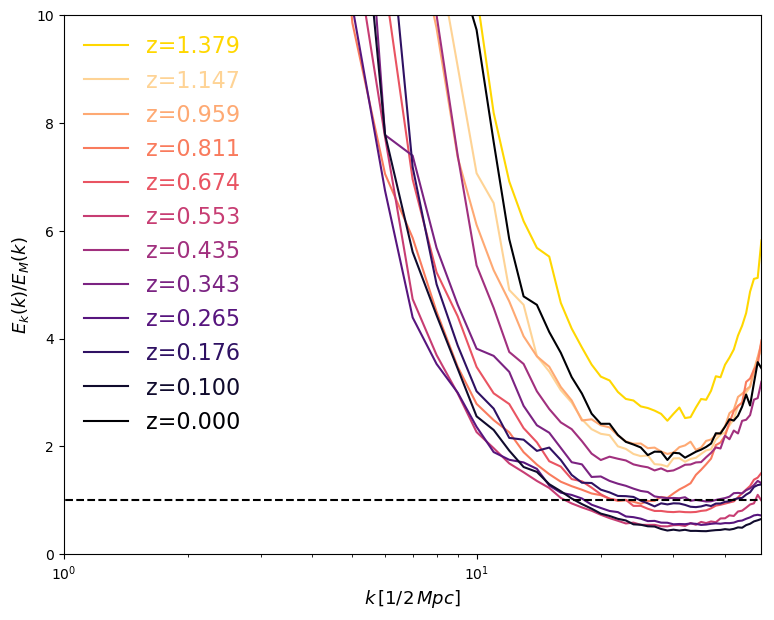}
    \caption{\textit{Top panel:} Evolution of the spectral kinetic and magnetic energy in the 
    simulation box of $100^3$ cells. The top spectra correspond to the kinetic 
    energy and the bottom spectra correspond to the magnetic energy. 
    The velocity
    power spectrum was multiplied by $\sqrt{n}$, $n$ being the gas density, in
    order for the  spectra to have the same units. \textit{Bottom panel:} Ratio of kinetic to magnetic
    energy as function of the wave number. The horizontal dashed line indicates where we have
    equipartition.}
    \label{fig04:Spec_evolution_all}
\end{figure}

In this section we focus on the spectral features of the magnetic and kinetic energy power spectra, whose entire evolution is given in 
Fig. \ref{fig04:Spec_evolution_all}.  The spectra are shown using comoving units and are computed within  a box of comoving size
$L=1.58$ Mpc, which moves with with the cluster centre 
identified as described in Section \ref{section:E5A_evolution}.
The first thing to notice is that the shape of both spectra change very little over the period from
 $z=1.379$ to $z=0$,  while the normalization increases whenever a minor merger occurs.
In the bottom panel of Fig. \ref{fig04:Spec_evolution_all}, we
can observe that the evolution of the magnetic energy spectra shows a global increment on the magnetic field strength up to an approximate state of equipartition for 
$k \sim 20-50$ (corresponding to scales $\sim 30-80$ kpc). 
This means that during a significant fraction
of the system evolution the  magnetic tension is strong enough
to prevent the further bending of the magnetic lines, as 
would be expected from 
a classic small-scale dynamo. However, the range of scales in which equipartition is reached does not evolve monotonically with time (as expected in a classic dynamo), but it fluctuates in time, with features that are non-trivial to isolate. In particular,
the epochs where there is no equipartition coincide with
the occurrence of mergers, i.e. when the cluster is more 
perturbed.
The various kinetic injections driving turbulence in this system will continuously change the magnetic field topology on 
spatial scales larger than the equipartition scale. 

\begin{figure}
	\includegraphics[width=\columnwidth]{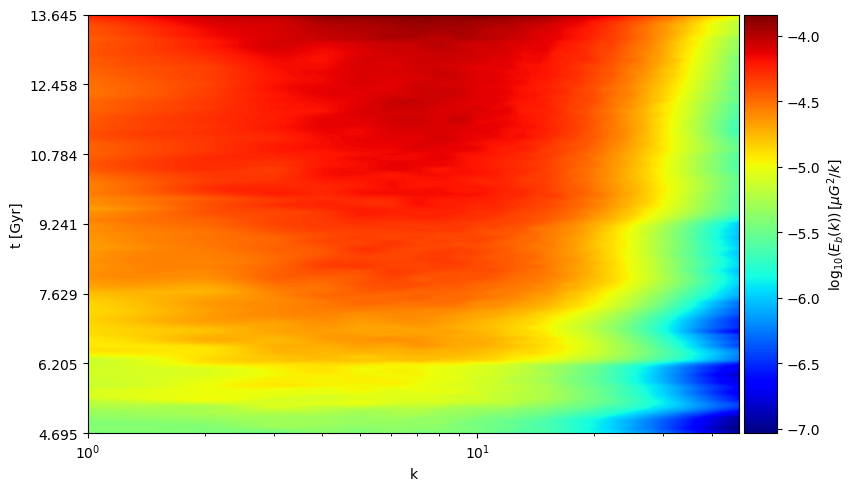}
	\includegraphics[width=\columnwidth]{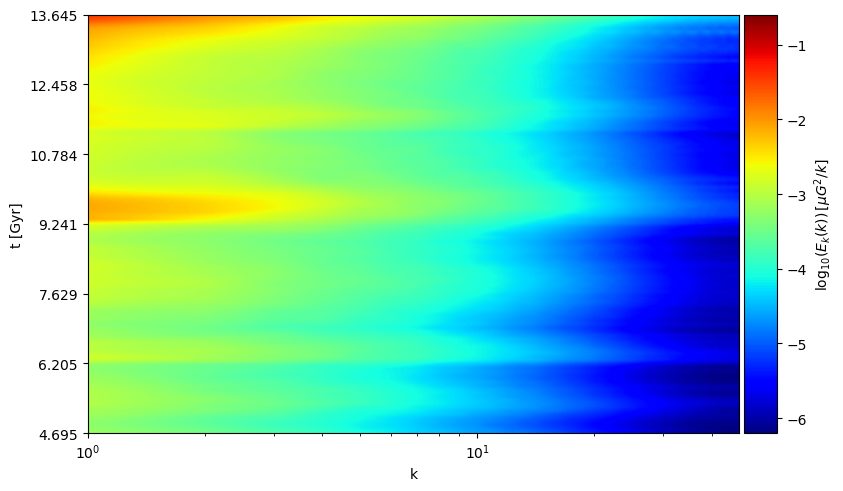}\\
    \caption{Evolution of the spectral energy in the 
    simulation box of $100^3$ cells. The \textit{top panel} shows the
    corresponding evolution of the magnetic energy and the \textit{bottom
    panel} shows the evolution of the kinetic energy.}
    \label{fig04:Spec_evolution}
\end{figure}

In order to examine the evolution of E5A, in Fig.~\ref{fig04:Spec_evolution} we colour coded the amplitude of the magnetic and kinetic spectra as a function
of time. This spectral time sequence shows the entire evolution of the ICM as a function of time and spatial scale. 
 As the system evolves, the magnetic power increases and tends to shift towards smaller scales, while the kinetic spectrum is always characterized by a maximum at $k =1$, which mirrors the fact that the forcing of turbulence always occurs on scales $\geq 1 ~\rm  Mpc$. 
 Merger events can be seen as horizontal stripes in the plot, which correspond to the injection of kinetic energy.
  
\begin{figure}
	\includegraphics[width=\columnwidth]{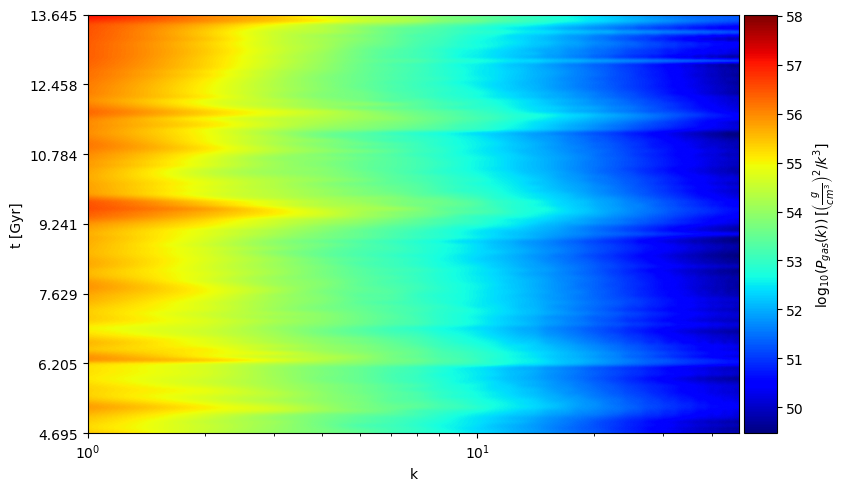}
    \caption{Density power spectrum as a sequence of time.}
    \label{fig05:PS_evolution}
\end{figure}

 The resulting amplification of the magnetic field strength is then a complex interplay between compression and the small-scale dynamo.
This is best shown by the  appearance of dense gas structures at a similar time, as shown in the power spectra of gas density in  Fig. \ref{fig05:PS_evolution}, which is consistent with the 
relation between velocity and density fluctuations in the 
stratified ICM \citep[e.g.][]{2014A&A...569A..67G}.   
A general trend is that every
merger shifts the magnetic spectral power towards
{\it higher} wave numbers, i.e. during most of these events the peak of the magnetic energy spectrum moves towards smaller spatial scales, unlike what is expected from the standard dynamo model, and most likely due to gas compression.  
As cluster mergers generate shocks and bulk flows that
enhance the gas density and compress the magnetic field lines, this can also increase the
normalization of the spectrum. Furthermore, it can also move the peak of the spectrum to higher wave numbers because the magnetic field lines get stretched along the merger direction. 

Simultaneously,  mergers inject turbulence, and only after the latter has decayed to small scales (where the eddy turnover time is the shortest), 
the peak magnetic spectra shifts towards lower wave numbers and the magnetic field is boosted again. This effect is characteristic
of a small-scale dynamo.

Our analysis implies that both, compressive and dynamo amplification, 
tend to be present at the same time in galaxy clusters. This causes a difficult evolutionary pattern in the simulated ICM, 
 adding complexity to what has been previously obtained by more idealized MHD simulations \citep[e.g.][]{2016ApJ...817..127B,2015Natur.523...59M}. 


\begin{figure}
	\includegraphics[width=\columnwidth]{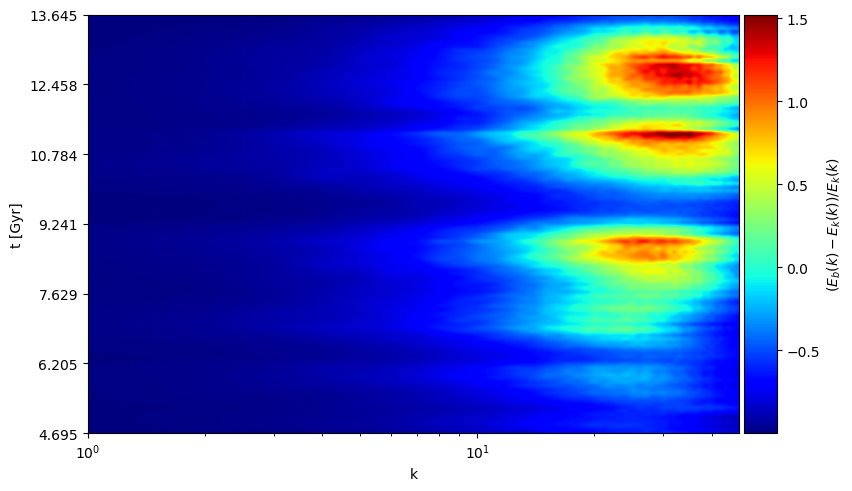}
    \caption{Energy residual evolution corresponding to
    the energies in Fig.\ref{fig04:Spec_evolution}. The highest
    values appear at small scales showing how the amplified
    magnetic field is able to overcome the kinetic pressure.
    }
    \label{fig06:Res_evolution}
\end{figure}

For better visualization, Fig. \ref{fig06:Res_evolution}
shows the residual between magnetic and kinetic spectral energies also as a spectral time sequence plot.  At all epochs, the excess magnetic energy is found on wave numbers
$k > 10$ (corresponding to scales $<160$ kpc), showing that
after merger events 
the magnetic tension gets strong enough  to overcome further
bending of the magnetic lines only at small scales.
The magnetic amplification starts \textit{only} after merger events
because the turbulence injected takes a few eddy-turnover times
to cascade. 

In fact, if the kinetic energy injection is high enough, as we can
observe around $t\sim 9.8$ Gyr in Fig. \ref{fig06:Res_evolution}, the amplification is slowed down. 

\begin{figure}
	\includegraphics[width=\columnwidth]{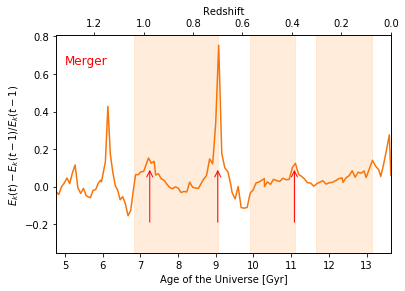}
    \caption{Kinetic energy residual as function of time.
    The red arrows are related
    to the time when an in-falling gas clump crosses the centre
    of the cluster. The
    shaded areas are identified directly with Fig.\ref{fig06:Res_evolution}, therefore indicating
    the periods of amplification.}
    \label{fig04:Residual_evolution}
\end{figure}

In order to identify the specific times of kinetic energy injection, we
plotted in Fig. \ref{fig04:Residual_evolution} the difference of the total kinetic
energy in the simulation box at timestep $t_i$ with respect to the previous timestep, $t_{i-1}$. 
A peak in this plot can account mainly for either the entrance of a clump into the
simulation box, a shock traveling across the cluster or a reflected shock. Since
we are interested in studying the amplification periods identified in Fig. 
\ref{fig06:Res_evolution}, we restrict ourselves to point out only some
of these events confirmed by visual inspection 
with red arrows in Fig. \ref{fig04:Residual_evolution}. The
shaded areas in the plot are placed as a reference for the amplification phases
found in the spectral time sequence of Fig. \ref{fig06:Res_evolution}. 
We noticed that, the maximum kinetic injection appears to happen either when gas substructures
cross close to the cluster centre,  which typically leads to shock waves ($\mathcal{M} \sim$ 2--3 in this case, as we measured with a velocity-based shock finder following \citealt{va17turbo}) sweeping through the cluster; 
or when there is a continuous
injection of turbulence by minor mergers (period between $t \sim$ 12--13 Gyr). 
In the first case, the most significant boosts of kinetic energy are followed by the compression of the magnetic field spectra.
The injection of large amounts of kinetic energy on large scales
impact the magnetic field only after $\lesssim 1$ Gyr (white areas after first and second red arrows in Fig. \ref{fig06:Res_evolution}), suggesting that a small-scale
dynamo is activated only after such amount of time. 
In the second case, continuous minor mergers contribute to 
the magnetic amplification at small scales by
starting to shift the power towards higher scales (period between $t \sim$ 12--13 Gyr).
This seems to suggest that minor mergers significantly power the
 small-scale dynamo amplification.

\begin{figure}
    \centering
    \includegraphics[width=\columnwidth]{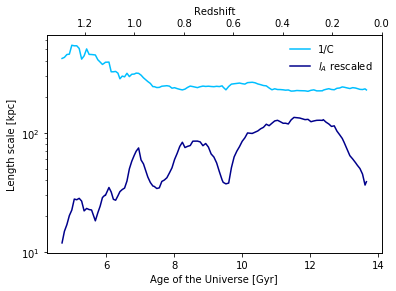}
    \caption{Evolution of the MHD scale and the outer scale of
    the magnetic spectrum (inverse of the C parameter). Note that the MHD scale is rescaled by a factor of 50 for ease of comparison.}
    \label{fig:mhd}
\end{figure}

Finally, we studied the evolution of the MHD scale ($l_A$) using the result from \citet[][]{brunetti2007}:

\begin{equation*}
\resizebox{.96\hsize}{!}{$
    l_A \sim 3 \left(\frac{B}{\mu G} \right)^3
    \left( \frac{L_0}{1 \, \textnormal{Mpc}}\right)
    \left( \frac{\sigma_{v}}{10^3 \textnormal{km} \textnormal{s}^{-1}} \right)^{-3}
    \left( \frac{n}{10^{-3} \textnormal{cm}^{-3}}\right)^{-3/2} \, \textnormal{kpc},
    $}
\end{equation*}

where $L_0$ is the reference scale within the Kolmogorov inertial range and $\sigma_{v}$ is the rms velocity within the scale $L_0$.
In this case, we measure the turbulent velocity by filtering the large motions on $\approx 300$ kpc.
We obtain a distribution of the MHD scale for all of our snapshots and select the mean at
each time. In
Fig. \ref{fig:mhd} we  show the resulting evolution of the corresponding scale ($l_A$)
and compare it to the evolution of the outer scale of the magnetic spectrum ($1/C$). It has been
suggested in former studies \citep[e.g.][]{2016ApJ...817..127B,2015Natur.523...59M} that
$l_A$ will follow closely the evolution of the outer scale of the magnetic spectrum. Our analysis suggests that in reality the evolution of magnetic fields during
mergers is more complicated than that. The system is significantly affected by compression and large-scale coherent motions, whose energy is larger than the small-scale turbulent energy on $\leq 300 ~\rm kpc$ scales. 
 In fact, the injected energy may contribute to advect magnetic field lines on large scales
($>100$ kpc). Overall, this means that our galaxy clusters exhibit cumulative turbulence cascades with different injection timescales, able to amplify the existing magnetic
fields via a dynamo action. 
Under these conditions, the evolution of the outer scale is mismatched with respect to that of the MHD scale.
This has important implications for the future surveys of magnetic fields in galaxy clusters. 
The interpretation of  magnetic field spectra inferred by Faraday Rotation will not uniquely constrain the 
magnetic amplification coming from a small-scale dynamo,
but may also be contaminated by compression amplification
coming from large-scale gas flows.

\subsubsection{Evolution of best-fit parameters for cluster E5A}

\begin{figure}
    \centering
    \includegraphics[width=\columnwidth]{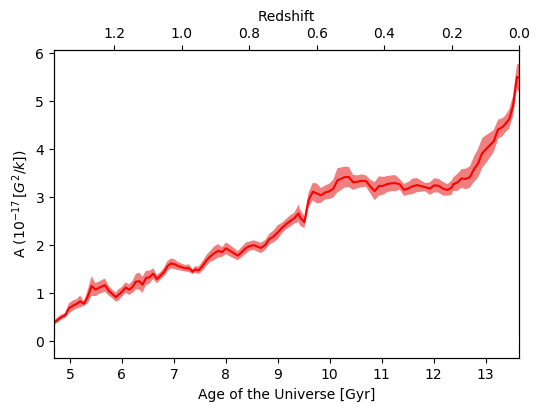}\\
    \includegraphics[width=\columnwidth]{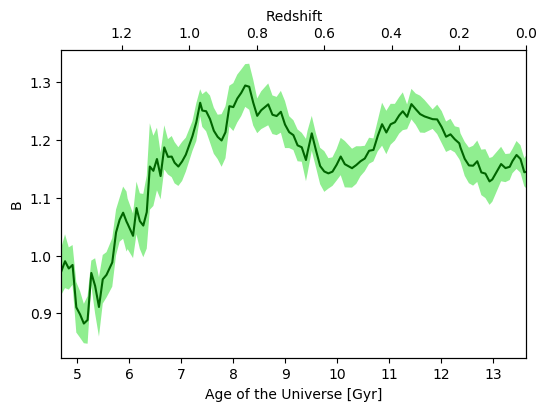}\\
    \includegraphics[width=\columnwidth]{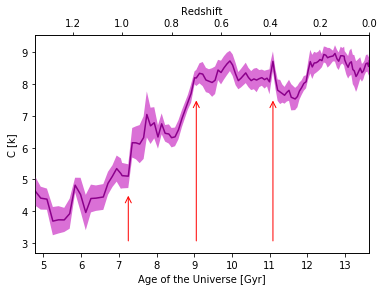}
    \caption{Evolution of the best-fit parameters $A$, $B$ and $C$ obtained by means of Eq. (\ref{fit_eq}). The $2\sigma$ error envelopes are shown in lighter shades.}
    \label{fig08:Params_evolution}
\end{figure}

Following the same approach of   Section \ref{section:fit}, we 
proceeded with the fitting of all magnetic spectra in the evolution of E5A,
which yields the evolutionary tracks shown in  Fig. \ref{fig08:Params_evolution}. The top
panel shows the normalization of the magnetic energy spectrum, where
we can see a clear result: the overall amplification of the magnetic
field continues to grow but steepens more where mergers occur. In fact, we observe
that the normalization almost increases by one factor on the last $\sim 0.5$ Gyr where
a major merger is about to happen.
As a consequence of these events and the other effects previously mentioned, the magnetic growth is not linear.
While the total magnetic energy increases by a factor of $\sim$
40-50 (as mentioned in Section \ref{section:E5A_evolution}), the
normalization of the spectrum only increased by a factor of $\sim 5$
in nearly 9 Gyr.

In the middle and bottom 
panels of Fig. \ref{fig08:Params_evolution} we show the evolution
of the parameters B and C. It is notable that both evolution patterns seem to be correlated. The evolution of C 
(wave number corresponding to the outer scale of the magnetic spectrum)
also shows a correlation with some identified merger events: the red arrows
over-plotted corresponding to those in Fig. \ref{fig04:Residual_evolution}.
Mergers induce an immediate change of the outer scale of the spectrum by shifting the power towards smaller scales. While this pattern is less obvious in the evolution of the parameter B, we can observe that mergers also induce an immediate broadening of the spectrum. These combined effects can be directly associated with the action of compression. A
particular thing to notice is that, the change on B and C at the last (third arrow) merger event is not as large as the previous events. This
suggests that at this point, the cluster has had enough turbulence input
(at different injection scales and timescales) to amplify the magnetic
field at smaller scales, making it harder for the spectrum to broaden or
shift its power to even smaller scales.

%

\section{Numerical aspects}\label{section:numerics}

As in Paper I, we relied on the Dedner cleaning algorithm \citep[][]{ded02} to get rid of magnetic monopoles. The main limitation of this method is the reduction of the effective dynamical range, compared to  Constrained Transport (CT) schemes at the same resolution, due to the intrinsic dissipation of the scheme by  $\nabla \cdot \vec{B}$ cleaning waves which keep the numerical divergence under control \citep[][]{kri11}. 
Several groups have tested that the Dedner cleaning  method is robust and accurate for most idealized test problems, as long as the resolution is opportunely increased \citep[e.g.][]{wa09,wang10,enzo14}. Even in the test of more realistic astrophysical applications, the Dedner method has been shown to quickly converge to the right solution, unlike different approaches to clean $\nabla\cdot \vec{B}$ preserving  at run time
\citep[][]{2013MNRAS.428...13S,2016MNRAS.455...51H,2016MNRAS.461.1260T,2018MNRAS.476.2890B}. 

 Despite the numerical dissipation introduced by the Dedner cleaning, all important features discussed in this paper (e.g. the peak in the power spectrum of magnetic fields, and the equipartition scales) are much larger than the length scales affected by numerical dissipation: e.g. the peak of power spectra are typically on scales $\sim 25-50$ larger than the minimum cell size in our the simulation. While the dissipation in the Dedner scheme can considerably slow down the first stage of the dynamo amplification \citep[][]{2016ApJ...817..127B}, once that magnetic structure becomes sufficiently large, they are relatively unaffected by numerical dissipation.

In Paper I we verified that in the largest part of the simulation box, the numerical divergence of $B$ is of order $\sim 2\text{-}3 \%$ of the local magnetic field strength, i.e.   $\leq 10^{-4}$ of the magnetic energy on larger scales.
We refer the reader to the recent review by \citet{review_dynamo} for a broader discussion of
the resolution and accuracy of different MHD schemes in the context of small-scale dynamo processes in galaxy clusters.\\

Our simulations neglect physical processes other than gravity and (magneto)hydrodynamics, 
such as  radiative gas cooling, chemical evolution, star formation and feedback from active galactic nuclei.  In this way, we can more easily isolate the effects of compression and dynamo from additional amplification caused by feedback and gas overcooling.\footnote{See however \citet{2018arXiv181008619K}, for a possible way of monitoring the growth of different magnetic field components within the same simulation.}
Comparisons between the predictions of primordial and astrophysical seeding scenarios of magnetic fields with {\enzo} can be found in \citet{va17cqg}. For recent high-resolution simulation of extragalactic magnetic fields with a moving-mesh algorithm we refer the reader to \citet{ma18} and to the recent review by \citet{review_dynamo}.\\

While the initial {\it topology} of possible seed magnetic fields is unknown, we tested in Paper I that variations of the assumed initial topology of seed fields do not to affect the strength of simulated magnetic fields in the ICM at low redshift.  Variations of the assumed initial {\it strength} of magnetic seed fields are harder to test, as for very small seed fields resolving the Alfvenic scale $l_A$ becomes prohibitive and the amplification is stuck in the exponential regime for the entire cluster evolution \citep[][]{2016ApJ...817..127B}. In Paper I, we provided evidence that our simulated magnetic fields are fairly independent on the initial field strength only for
 $\geq 0.03 ~\rm nG$ (comoving) fields.  Future re-simulations at even higher resolution, or with less diffusive MHD schemes will be needed to test the scenario for lower seed fields.\\

Finally, as customary in simulations without explicit viscosity and resistivity, the numerical viscosity and resistivity are of the same order, meaning that the magnetic Prandtl number is $P_M=R_{\rm M}/R_{\rm e}=\nu/\eta \approx 1$.This assumption is reasonable enough given the existing uncertainties and the difficulties in the characterization of the magnetised plasma in galaxy clusters \citep[e.g.][]{2004ApJ...612..276S,bl11b,2016ApJ...817..127B}, and it further allows us to easily compare with the standard literature of small-scale dynamo in a box \citep[e.g.][]{cho14,2015ApJ...810...93P}. A few groups have explored the role of non-ideal MHD effects in cosmological simulations, such as the presence of a physical resistivity \citep[e.g.][]{bo11b,2018MNRAS.476.2476M}, whose usefulness to explain observed ICM magnetic fields has been recently questioned by new simulations \citep[][]{2018MNRAS.476.2890B}. 

\section{Summary and conclusions}
\label{sec:conclusions}

In this paper, we have presented new high-resolution cosmological MHD simulations of a sample of galaxy clusters, which allow us to study the 
spectral properties of magnetic amplification with unprecedented spatial and temporal detail.

In agreement with our earlier work, we find that we can reproduce cluster magnetic fields of the order of $\sim 1 -3 ~\rm \mu G$ with primordial fields of  $10^{-10}$ G (comoving) at  $z=30$.

We computed the magnetic energy spectra at $z=0$ for all the clusters in the sample.
The spectral shape remains similar across clusters, despite of their different dynamical states. 
We parameterize the magnetic spectra of all the
clusters in our sample at $z=0$ and as a function of time for the
merging cluster E5A by means of Eq. (\ref{fit_eq}). The
resulting best-fit parameters are used to characterize
 the magnetic properties of the ICM. 
In general, we could not find a simple one-to-one relation between the kinetic and magnetic spectra and the dynamical state of the clusters:  this indicates that highly perturbed systems, exhibiting more turbulence, do not necessarily imply higher values of the magnetic fields, and that the cycle of amplification of magnetic fields in the realistic ICM is complex.

The  normalization of the magnetic spectrum ($A$),
the spectrum width ($B$) and the inverse
of the outer scale of the spectrum ($C$) show
a positive correlation with the virial mass of each cluster. 
In addition, 
$B$ is correlated with the dynamical state of clusters. In general, we observe that the magnetic growth rate is larger for merging systems, 
while it is smaller in the relaxed system in our sample.

Finally, the outer scale of the magnetic spectrum ($\propto 1/C$) also correlates with the dynamical state of the
cluster: the relaxed system in our sample reaches higher values of the outer scale ($\sim 300$ kpc) compared
to merging ($\sim 230$ kpc) and post-merging ($\sim 200$ kpc) systems, possibly indicating that the dynamo has acted for a longer time in such systems.  We caution that the ubiquitous presence of large-scale bulk motions in the ICM may introduce larger correlation scales in the magnetic field, so that our best-fit parameters do not show an evident correlation with the last major merger of each cluster. This suggests that the history of minor mergers matters, but larger  statistics of simulated clusters would be necessary to reach firmer conclusions. 

Moreover, we studied the co-evolution of magnetic fields and the ICM properties in a merging cluster (E5A), which we could sample with a high time resolution. 
Our analysis reveals that the peak of the magnetic power spectrum shifts towards smaller spatial scales shortly after mergers, while overall it shifts to larger scales. In the cluster E5A, the peak of the magnetic power spectrum extends to $\sim 280$ kpc after $\sim 9$ Gyr of evolution, with equipartition at scales  $<160$ kpc. Large amounts of kinetic energy are injected by substructures that fall through the cluster which first amplify the magnetic field mainly via compression. These mergers prevent equipartition on the smallest scales, i.e.
when the cluster is more perturbed, equipartition is not 
reached at scales above our current resolution.

In the course of a merger, the spectrum broadens and the outer scale is shifted 
towards smaller scales. 
While we observe that the total magnetic energy is continuously
growing, the magnetic amplification at smaller scales starts only
after the mergers. This behaviour is driven by two mechanisms:
1) strong mergers introduce additional turbulence into the system that raises the kinetic energy above equipartition with the magnetic field. Nevertheless, this new energy will only become available 
for magnetic amplification after a few eddy-turnover times when the turbulence has already cascaded down to the smaller scales; 
Consequently, this changes the growth timescales by slowing
down the process of amplification soon after a merger event.
In particular, when there is a large input of kinetic energy, the magnetic amplification at small scales sets in only after $\sim 1~\rm Gyr$ since the merger.
\\

Finally, our work has important implications for the interpretation of existing or future radio observations of magnetic fields in galaxy clusters.
The total rotation measure $|RM|$ from  clusters is expected to scale $ \propto A/C $. 
 Therefore, our previous results imply that the RM only weakly depends on the  mass of the galaxy cluster.  We  measure a scatter of up to a  $\sim 4$  difference in RM between clusters of the same mass, while systems with a $\sim 2$ difference in mass can have the same RM, due to differences in their magnetic field correlation scale.  
 This implies that the RM across the cluster population probably is not universal, but can  significantly be affected by the complex sequence of amplification events in the past lifetime of each cluster, with important consequences in the predictions of the RM from galaxy clusters which should be observable by future radio polarisation surveys \citep[e.g.][]{2015aska.confE.105G,2015arXiv150102298T}. We defer this analysis to future work.

\section{Acknowledgements}
We acknowledge our anonymous reviewer for helpful comments on the first version of this manuscript. 
The cosmological simulations were performed with the {\enzo} code (http://enzo-project.org), which is the product of a collaborative effort of scientists at many universities and national laboratories. We gratefully acknowledge the {\enzo} development group for providing extremely helpful and well-maintained on-line documentation and tutorials. The analysis presented in this work made use of computational resources on the JURECA
cluster at the at the Juelich Supercomputing Centre (JSC), under
projects no. 11823, 10755 and 9016 and HHH42, and partially on the Piz-Daint supercluster
at CSCS-ETHZ (Lugano, Switzerland) under project s805. 
The original simulations on which this work is based have been produced by F.V. as PI on project HHH42 on JSC. 
We also acknowledge the usage of online storage tools kindly provided by the Inaf Astronomica Archive (IA2) initiave (http://www.ia2.inaf.it).   \\
P.D.F and F.V. and acknowledges financial support from the European Union's Horizon 2020 program under the ERC Starting Grant "MAGCOW", no. 714196. We acknowledge useful scientific discussions with K. Dolag, A. Beresnyak, J. Donnert, D. Ryu and T. Jones. 

\bibliographystyle{mnras}
\bibliography{paola,franco}

\appendix

\section{Earlier results on dynamo amplification in simulated clusters.}
\label{appen_2}

\begin{figure}
    \centering
    \includegraphics[width=\columnwidth]{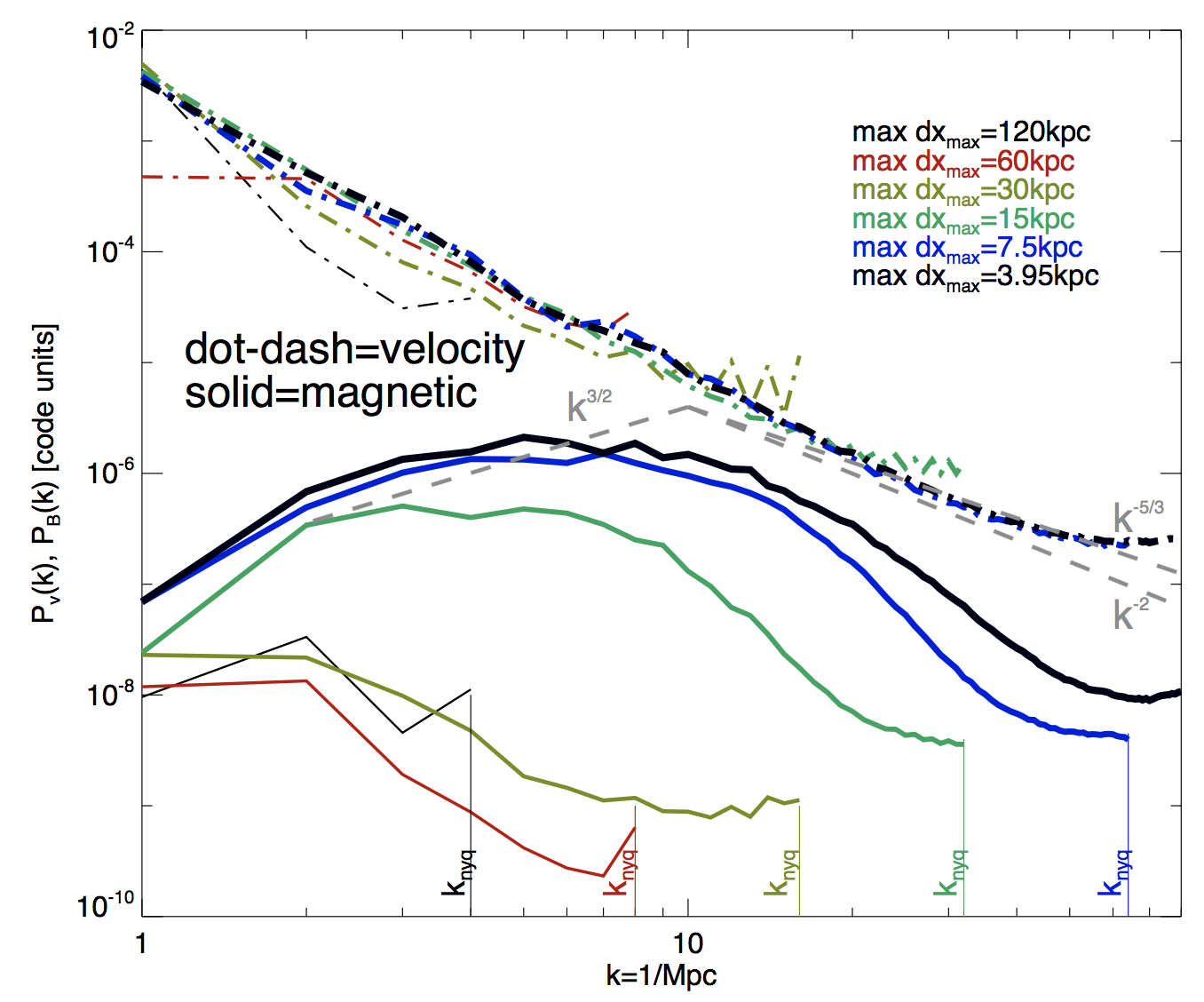}
    \caption{Power spectra of velocity (top lines) and magnetic field (lower curved lines) for resimulation of increasing resolution, presented in Paper I. The spectra are measured within the innermost $2^3 \rm Mpc^3$ of a simulated cluster at $z=0$, and clearly shows that the increase of resolution leads to an increase of the dynamical range (also marked by the sequence of Nyquist frequencies at the bottom of the panel) and results into a radial change in the magnetic spectrum for $\leq 16 \rm ~kpc$ resolutions.}
    \label{fig:spec_res}
\end{figure}

We summarize here the main results of our previous work \citep[][, Paper I in this work]{va18mhd}, which motivates the analysis performed in this paper.
Using an AMR prescription to refine most of the innermost regions of galaxy clusters with the {\enzo} code, in  \citet[][]{va18mhd} we simulated the growth  of 0.1 nG (comoving) magnetic field seed, assumed of a cosmological origin, from $z=30$ to $z=0$.
We simulated the field growth as a function of the maximum cell resolution for a  Coma-like galaxy cluster ($\sim 10^{15} M_{\odot}$) an starting from the same initial field, and observed the onset of significant small-scale dynamo for resolutions $\leq 16 ~\rm kpc$, with near-equipartition magnetic fields on $\leq$ 100 kpc scales.
For the  best resolved run ($\approx 4 ~\rm kpc/cell$), we measured  a final magnetic fields strength of  $\sim 1-2~ \rm \mu G$ in the cluster core, with a radial profile that scales as $B(n) \propto n^{0.487}$ (where $n$ is the gas density). 
For lower resolution, the magnetic field gets increasingly smaller, with a flatter radial profiles and a magnetic power spectrum of a power-law shape. 
In summary, the following are the key evidences that support that our runs do feature a resolved small-scale dynamo:
\begin{itemize}
    \item the measured dependence of magnetic field strength and the effective resolution of the simulation: only when the numerical Reynolds number exceeds $R_e \sim 10^2$ the magnetic field reaches values much larger than what gas compression ($\propto n^{2/3}$) can produce;
   \item the onset of the curved magnetic field power spectrum only when the spatial resolution exceeds a critical value (estimated to be $\sim 16 \rm ~kpc/cell$, even if this may vary with the adopted numerical scheme, e.g. \citealt{review_dynamo}), indicating that only at a large enough Reynolds number and high enough resolution we have enough solenoidal turbulence and we can resolve the $l_A$ scale (Fig.\ref{fig:spec_res});
   \item the slope of the power spectra for low wavenumbers is compatible with the Kasantsev model of dynamo $P_B \propto k^{3/2}$ \citep[e.g.][]{2004ApJ...612..276S},  while after the peak the spectrum rapidly steepens from $\propto k^{-5/3}$ to $\propto k^{-2}$ or less, consistent with \citep[e.g.][]{2015ApJ...810...93P,2017arXiv170405845R};
   \item the evolution of magnetic fields in our most resolved simulation, and its relation with the measured dissipation of kinetic turbulent energy, which indicate a $\sim 4\%$ dissipation rate of turbulent into magnetic energy, in line with \citet{2015Natur.523...59M} and \citet{2016ApJ...817..127B};
   \item the measured  anti-correlation between the curvature of magnetic field lines in our most resolved simulation and the magnetic field strength, as expected in the dynamo regime \citep[e.g.][]{2004ApJ...612..276S};
    \item the measurement that the $l_A$ scale, estimated following in \citet{bl07}, which is well resolved for a good fraction of our cluster volume;
    \item the independence of the magnetic profile and power spectra at $z=0$, for $\geq 0.03 \rm ~nG$ (comoving), above which our setup ensures to resolve $l_A$ in a large fraction of the cluster volume.
\end{itemize}

Moreover,  the topology of the magnetic fields at $z=0$ produces profiles of Faraday Rotation of background polarised sources in good agreement with the real observations of the Coma cluster, which are the most stringent to date  \citep[][]{bo10,bo13}. 
A significant new finding of our first analysis in Paper I is also the detection of a significant non-Gaussian distribution of magnetic field components in the final cluster, which results from the superposition of different amplification patches mixing in the ICM.  

All results obtained from this first study are also confirmed with the larger set of cluster simulations
which is object of this paper. 

\section{Correlating the best-fit parameters}
\label{appen_corr}

\begin{figure}
    \centering
    \includegraphics[width=\columnwidth]{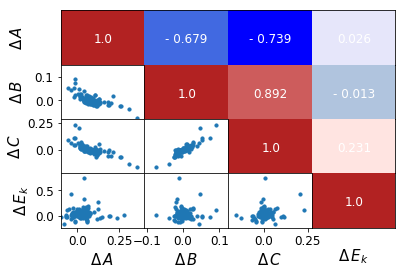}
    \caption{Cross-correlation matrix of the best-fit parameters and
    the kinetic energy changes in the system. The Pearson correlation
    coefficients are indicated in the upper part of the diagonal and the corresponding
    scatter plots are shown in the lower part of the diagonal.}
    \label{fig:matrix}
\end{figure}

We computed the cross-correlation
matrix of the change in time of the best-fit parameters $A$, $B$, $C$ and the kinetic energy $E_k$ and
show the result in Fig.\ref{fig:matrix}. Where $\Delta$ of a variable
$Q$, is defined as $(Q(t)-Q(t\text{-}1))/Q(t\text{-}1)$ as in 
Fig. \ref{fig04:Residual_evolution}. The Pearson coefficients
for all the cross-correlations are shown in the upper part
of the diagonal in Fig. \ref{fig:matrix}. In this way, we can better quantify
the existing correlations and interpret them:

\begin{enumerate}
    \item $corr(\Delta A, \Delta B)$: 
    a positive change in the normalization
    implies a negative change in the parameter B. 
    This implies that a sudden increment on the normalization narrows down
    the spectrum width shortening the magnetic growth timescale.
    Therefore, the growth rate increases over time.
    \item $corr(\Delta A, \Delta C)$: an increment in the normalization
    implies that $C(t)<C(t\text{-}1)$, i.e the power is shifted towards larger scales. 
    We attribute this feature to the presence
    of dynamo amplification. This conclusion is supported
    by Section \ref{sec:spectral_evolution}, where every merger
    event carrying enough kinetic energy was shown to shift the 
    magnetic spectrum towards smaller scales (i.e. amplifying via compression).
    
    \item $corr(\Delta B, \Delta C)$: 
    a wider spectrum coming along with a shift of
    the outer scale towards smaller scales
    is directly related to the action of compression. This matches our previous interpretation of Fig. \ref{fig06:Res_evolution}, where compression shifts the power to smaller scales 
    (i.e. $C(t\text{-}1)<C(t)$) 
     and the new turbulent cascade does not play a role in the amplification instantaneously, but after an eddy-turnover time.

    \item $corr(\Delta C, \Delta E_k)$: a shift of the outer scale 
    towards small scales is weakly correlated with the injection of kinetic energy.
    In this case, we checked the cross-correlation at each time and identified
    the times corresponding to some merger events (red arrows in Figs.\ref{fig04:Residual_evolution} and \ref{fig08:Params_evolution}). The first two arrows corresponding
    to small clumps falling into the cluster show a higher correlation than the last arrow which corresponds to a larger clump.  
    A plausible explanation is that the first two events generated sufficient turbulence that allowed the magnetic field to
    grow also at smaller scales (via the small-scale dynamo), so by the time of the third event,
    the effect of compression is not enough to shift the outer scale towards smaller scales anymore. This can be considered as a momentary state of "balance" between the dynamo and compression effects and it would also explain the period of amplification between $t\sim$ 12-13 Gyr.
\end{enumerate}

\end{document}